\documentstyle[aps,prl,twocolumn,floats,psfig,epsf]{revtex}

\begin{document}
\draft
\twocolumn[\hsize\textwidth\columnwidth\hsize\csname@twocolumnfalse\endcsname
\title{Generalized Crossover in multiparameter Hamiltonians}

\author{Pietro Parruccini and Paolo Rossi}
\address{Dipartimento di Fisica dell'Universit\`a di Pisa and INFN-Sezione di Pisa\\
Via Buonarroti, 2 - 56127 Pisa, Italy.\\
{\bf e-mail: \rm
{\tt parrucci@df.unipi.it},
 {\tt rossi@df.unipi.it}
}}

\date{\today}

\maketitle

\begin{abstract}
Many systems near criticality can be described by Hamiltonians involving several relevant couplings and possessing many nontrivial fixed points. A simple and physically appealing characterization of the crossover lines and surfaces connecting different nontrivial fixed points is presented. Generalized crossover is related to the vanishing of the RG function $Z_t^{-1}$.

An explicit example is discussed in detail based on the tetragonal GLW Hamiltonian.
\end{abstract}
\pacs{PACS Numbers: 05.10.Cc, 05.70.Fh, 75.10.Hk, 75.10.-b}
\vspace{0.5cm}
]

% ========================= BODY =========================
%\narrowtext

According to Wilson Renormalization Group (WRG) theory, the critical properties of physical systems undergoing second order phase transitions are well described by the infrared behavior of (quantum) field theories belonging to the appropriate universality class.

In practice, however, it is usually quite hard to reproduce the experimental conditions corresponding to strict criticality and to verify the scaling predicted by (massless) field theory. Scaling is obscured by the influence of non-universal subleading contributions related to the presence of irrelevant operators in the effective Hamiltonian, but these effects are depressed in the proximity of a critical point and, under definite assumptions about the range of physical parameters explored, one may often show that a limited number of parameters is sufficient for a description of physics in the critical region.

In such situations the field theoretical approach is not helpless, and in particular one may show that, when the dynamics is dominated by the competition of a few fixed points of the RG trasformations, one may describe the behavior of the system in terms of ``effective''  exponents, whose values depend on the relative distance of the concrete physical situation from the competing fixed points. This dependence of the effective exponents from some nonuniversal parameter is usually termed ``crossover'', and it has been showed that crossover phenomena can be consistently studied in the context of (massive) field theory, and the functional dependence of the exponents is amenable to the coupling dependence of the standard RG functions.

Accurate theoretical and numerical studies of this phenomenon have been presented in the literature, mainly focussing however on the crossover between a trivial (Gaussian) fixed point and an attractive Wilson-Fisher point, in the presence of short or medium-range interactions.\cite{B-B-l,Fisher-prl,L-B-B-pre,L1,prv}
However there are physical situations characterized by the presence of a larger number of nontrivial fixed points. While only one among them is fully attractive and represents the physics of the second-order phase transition, the other nontrivial points exert some attraction on the RG trajectories, and as a consequence we may expect that, in the neighbourhood of criticality, the system be quite accurately described by points in the parameter space which lay near or above special RG trajectories connecting the different fixed points.
Generalized crossover exponents may be defined along these trajectories. Their values will in general interpolate between the values taken by the standard critical exponents at different critical points. In experimental measurements, under proper assumptions, it is reasonable to expect that sets of measured exponents will correspond to specific points along these curves.

It may therefore be useful to find intrinsic characterizations of these generalized crossover curves, which only in very simple and specific examples can be deduced directly from inspection of the relevant RG equations.

In order to study this problem, we found convenient to take a specific field-theoretical model, which was recently discussed in literature as the Tetragonal Landau-Ginzburg Wilson Hamiltonian (TLGW)\cite{Aharony-76,Pelissetto:2000ek}. We found that it has a sufficiently wide variety of fixed points to allow a rather general discussion of the issue at hand. The results we obtained in this specific example can be easily extended to many other systems where a similar multiplicity of non trivial fixed points is present.

Our starting point is the following Hamiltonian:
\begin{eqnarray}
{\mathcal{H}}[\phi]= \int d^dx  \left\{ \frac{1}{2}\sum_{i,a} \left[ {\partial}_{\mu}\phi_{a,i}^2(x)+r \phi_{a,i}^2(x) \right] \right. \nonumber\\
 \left. +\frac{1}{4!} \sum_{i,j,a,b}(u_0+v_0 \delta_{ij}+w_0 \delta_{ij} \delta_{ab}) \phi_{a,i}^2(x) \phi_{b,j}^2(x)\right\} 
\end{eqnarray}
where  $a,b=1,2,\dots M$ and $i,j=1,2,\dots N$.

The models with $M=2$ are physically interesting since they should describe the critical properties in some structural and antiferromagnetic phase transitions and they are sufficiently general for the purpose of illustrating our results. 

The RG functions $\beta_u$, $\beta_v$, $\beta_w$ and $\eta_{\phi}$, $\eta_{t}$
are known up to six loops, and it is possible to study the fixed points of the models and their stability properties by solving the equations for the common zeroes of the $\beta -$functions and evaluating the eigenvalues of the stability matrix.\cite{Pelissetto:2000ek}

The $\epsilon -$expansion analysis of the tetragonal Hamiltonian indicates the presence of eight fixed points. Not all of them however do actually  represent different indipendent physical situations, because of the symmetry
\begin{eqnarray} 
(u_0,v_0,w_0)\rightarrow(u_0,v_0+ \frac{3}{2}w_0,-w_0)
\end{eqnarray}
possessed by the above Hamiltonian in the case $M=2$.

One may verify that, while the four fixed points lying on the $w_0=0$ plane are really distinct, the two fixed points corresponding to $w_0<0$ can be obtained from the two points in the half space $w_0>0$ by a symmetry transformation. 
We may therefore restrict ourselves to $w_0 \geq 0$ with no loss of generality.

The six distinct fixed points can be classified according to their symmetry properties; with obvious notation we shall identify them by the following names:
\begin{eqnarray*}
&&G=Gauss \rightarrow (u_0=v_0=w_0=0) \\
&&I=Ising \rightarrow (u_0=v_0=0) \sim I' \\
&&H=Heisenberg \rightarrow (v_0=w_0=0)  \\
&&XY \rightarrow (u_0=w_0=0) \\
&&T=Tetragonal \rightarrow (w_0=0)  \\
&&C=Cubic \rightarrow (v_0=0) \sim C'
\end{eqnarray*}

The symmetry properties of the Hamiltonian reflect themselves into symmetries of the $\beta -$functions. These in turn imply the existence of subspaces of the parameter space ($u_0,v_0,w_0$) which are stable under RG transformations. One may easily show that, to all orders of perturbations theory, the following initial conditions are preserved by RG transformations:

\begin{itemize}
\item $u_0=0$: a plane including $G$, $I$, $I'$ and $ XY$. 
\item $w_0=0$: a plane including $G$, $H$, $XY$ and $T$. 
\item $v_0=0$: a plane including $G$, $H$, $I$ and $C$. 
\item $v_0+ \frac{3}{2} w_0=0$: a plane including $G$, $H$, $I'$ and $C'$.
\end{itemize}
An analysis of the stability matrix can be performed in full parameter space and in each of the invariant subspaces, leading to the following general conclusions:
\begin{itemize}
\item $G$ is completely unstable w.r. to any perturbation.
\item $H$, $I$ and $I'$ are attractive w.r. to the Gaussian point, else unstable w.r. to all perturbations.
\item $C$, $C'$ are stable in the subspaces $v_0=0$ and $v_0+ \frac{3}{2}w_0=0$ respectively, but their stability matrix possesses a negative eigenvalue in full parameter space.
\item $XY$ is certainly stable in the subspace $u_0=0$ and probably also in full parameter space, in which case $T$ has a direction of instability in the $w_0=0$ subspace, leading towards $XY$.\cite{Pelissetto:2000ek}
\end{itemize}

Most previous studies of crossover have been concerned with ``crossover lines'' connecting the Gaussian fixed point $G$ with nontrivial fixed points along RG trajectories. In the model at hand, the straight lines connecting $G$ to the points $I(I')$, $XY$ and $H$ are such crossover lines, and the corresponding crossover exponents can be easily related to the RG functions obtained by specializing the general expressions to the values taken along these lines:
\begin{eqnarray*}
\beta_I(w) \equiv \beta_w(0,0,w), && \qquad \eta_{t_I}(w)\equiv \eta_t(0,0,w), \\
\beta_{H}(u) \equiv \beta_u(u,0,0), && \qquad \eta_{t_{H}}(u)\equiv \eta_t(u,0,0), \\
\beta_{xy}(v) \equiv \beta_v(0,v,0), && \qquad \eta_{t_{xy}}(v)\equiv \eta_t(0,v,0) 
\end{eqnarray*}

In particular, the function $Z_t^{-1}$, related to the renormalization of the one-particle irreducible two-point function with an insertion of the operator $\sum_{i}\phi_{a,i}^2(x)$, can be evaluated along the crossover lines simply by integrating the corresponding differential equation
\begin{eqnarray}
\left[ \beta(z) \frac{\partial}{\partial z}+\eta_t(z) \right]Z_t^{-1}(z)=0
\end{eqnarray}
where $z$ is the generic coupling which parametrizes the crossover line.

It it relevant to our purposes to notice that, being $z^*$ the fixed point value of the coupling, such that $\beta(z^*)=0$, as a consequence of the above equation the function $Z_t^{-1}(z)$, under the ``nontriviality'' assumption $\eta_t(z^*)< 0$, enjoys the property $Z_t^{-1}(z^*)=0$. We may appreciate that other choices of the renormalization function $Z$, differing from $Z_t$ by powers of $Z_{\phi}$, will not alter our conclusion as long as the corresponding nontriviality condition $\eta(z^*)<0$ is satisfied.

In models characterized by a multidimensional parameter space, this notion of crossover must be supplemented with a description of the RG trajectories connecting different nontrivial fixed points.
As we shall immediately show, it is in general possible to define ``crossover surfaces'' in parameter space, enjoying the property that all the RG trajectories connecting nontrivial fixed points (and obviously the points themselves) lay upon these surfaces.

The formal proof of this statement for the above discussed tetragonal model goes as follows: we introduce the renormalization function $Z_t^{-1}(\bar{u},\bar{v},\bar{w})$ satisfying by definition the partial differential equation:
\begin{eqnarray}
\left[ \beta_{\bar{u}}\frac{\partial}{\partial{\bar{u}}}+\beta_{\bar{v}}\frac{\partial}{\partial{\bar{v}}}+\beta_{\bar{w}}\frac{\partial}{\partial{\bar{w}}}+ \eta_t \right]Z_t^{-1}(\bar{u},\bar{v},\bar{w})=0
\end{eqnarray}
with the boundary condition $Z_t^{-1}(0,0,0)=1$.

$Z_t^{-1}(\bar{u},\bar{v},\bar{w})$ obviously reduces to the above defined functions $Z_t^{-1}(z)$ whenever any two out the three couplings $\bar{u}$, $\bar{v}$, $\bar{w}$ are set equal to zero.

Let us now consider the two-dimensional surface identified by the condition:
$$Z_t^{-1}(\bar{u},\bar{v},\bar{w})=0$$
As a consequence of the differential equation obeyed by $Z_t^{-1}(\bar{u},\bar{v},\bar{w})$ and of the above condition, the vector field $\vec{\beta}\equiv\left[\beta_{\bar{u}}(\bar{u},\bar{v},\bar{w}), \beta_{\bar{v}}(\bar{u},\bar{v},\bar{w}),\beta_{\bar{w}}(\bar{u},\bar{v},\bar{w}) \right]$ is ortogonal to the vector field $\vec{\nabla}Z_t^{-1}\equiv\left[ \frac{\partial Z_t^{-1}}{\partial{\bar{u}}},\frac{\partial Z_t^{-1}}{\partial{\bar{v}}},\frac{\partial Z_t^{-1}}{\partial{\bar{w}}} \right]$ when the two vectors are evaluated at any point of the surface $Z_t^{-1}=0$, where $\vec{\beta}\cdot \vec{\nabla}Z_t^{-1}=0$

Therefore the RG trajectories going through any point of the surface $Z_t^{-1}=0$ are found to stay on the surface itself, since the local tangent to the trajectory, i.e. the vector field $\vec{\beta}$, is ortogonal to a vector normal to the surface (the gradient field $\vec{\nabla}Z_t^{-1}$). Our proof is now completed by the observation that all nontrivial points lay on the surface because, as previously discussed, they must satisfy the property $Z_t^{-1}(z^*)=0$.

An interesting consequence of our result is obtained by considering the intersections of the crossover surface $Z_t^{-1}(\bar{u},\bar{v},\bar{w})=0$ with the RG-stable planes obtained by setting $u_0=0$, $v_0=0$, $w_0=0$ and $v_0+ \frac{3}{2}w_0=0$ respectively. These intersections are obviously simple curves on the invariant planes, connecting couples of nontrivial fixed points and defining RG trjectories in the corresponding restricted parameter subspaces.

For the sake of definiteness let us consider the plane $u_0=0$: the function $Z_t^{-1}(0,\bar{v},\bar{w})$ satisfies the RG equation
\begin{eqnarray}
\left[ \beta_{\bar{v}}(\bar{v},\bar{w})\frac{\partial}{\partial{\bar{v}}}+\beta_{\bar{w}}(\bar{v},\bar{w})\frac{\partial}{\partial{\bar{w}}}+ \eta_t(\bar{v},\bar{w}) \right]Z_t^{-1}(\bar{v},\bar{w})=0.
\label{eq5}
\end{eqnarray}
It is straightfoward to show, by applying the implicit function theorem, that the function $\bar{v}(\bar{w})$ defined by the condition $Z_t^{-1}(\bar{v},\bar{w})=0$ satisfies the ordinary differential equation
\begin{eqnarray}
\frac{d\bar{v}}{d\bar{w}}= \frac{\beta_{\bar{v}}(\bar{v},\bar{w})}{\beta_{\bar{w}}(\bar{v},\bar{w})}
\end{eqnarray}
characterizing all RG trajectories in the $(\bar{v},\bar{w})$ plane, and furthermore it connects the unstable fixed points $I$, $I'$ to the stable point $XY$.

One cannot fail to notice that in deriving our result we only made use of very general properties of RG functions and equations.
Therefore we can draw the general conclusion that the condition $Z_t^{-1}=0$ may unambiguously characterize the ``crossover surface'' in wide classes of Hamiltonian systems involving many relevant parameters.

\appendix
\section{ The Large $\mathbf{N}$ Limit}
A rather explicit illustration of the mechanism described in the present paper is obtained by considering the tetragonal model in the limit of an infinite number of field components ($N \rightarrow \infty$). At variance with standard $O(N)$ vector models, the tetragonal model does not became trivial on this limit, because nontrivial contributions to all orders of $\bar{v}$ and $\bar{w}$ couplings are still present. However some semplifications occur which make our discussion, while stile quite general, formally much simpler.

In the large $N$ limit is possible to show that the RG functions take the following form:
\begin{eqnarray}
\beta_{\bar{u}}(\bar{u},\bar{v},\bar{w})&=&A(\bar{v},\bar{w})\cdot \bar{u}-B(\bar{v},\bar{w})\cdot \bar{u}^2 \\
\beta_{\bar{v}}(\bar{u},\bar{v},\bar{w})&=&\tilde{\beta}_{\bar{v}}(\bar{v},\bar{w})\\
\beta_{\bar{w}}(\bar{u},\bar{v},\bar{w})&=&\tilde{\beta}_{\bar{w}}(\bar{v},\bar{w})\\
\eta_{\phi}(\bar{u},\bar{v},\bar{w})&=&\tilde{\eta}_{\phi}(\bar{v},\bar{w})\\
\eta_{t}(\bar{u},\bar{v},\bar{w})&=&\tilde{\eta}_{t}(\bar{v},\bar{w})+B(\bar{v},\bar{w})\cdot \bar{u}
\end{eqnarray}

The system of equations $\tilde{\beta}_{\bar{v}}(\bar{v},\bar{w})=0$ and $\tilde{\beta}_{\bar{w}}(\bar{v},\bar{w})=0$ admits four sets of solutions $(\bar{v}^*,\bar{w}^*)$. For each set one finds two fixed points, corresponding to the values $\bar{u}^*=0$ and $\bar{u}^*= \displaystyle \frac{A(\bar{v}^*,\bar{w}^*)}{B(\bar{v}^*,\bar{w}^*)}$.

Because of the above relationships, the differential equation satisfied by the function $Z_t^{-1}(\bar{u},\bar{v},\bar{w})$ can be solved in the large $N$ limit by the Ansatz

\begin{eqnarray}
Z_t^{-1}(\bar{u},\bar{v},\bar{w})= \tilde{Z}_t^{-1}(\bar{v},\bar{w})\left[ 1- \bar{u} \cdot Y(\bar{v},\bar{w}) \right]
\end{eqnarray}

leading to the equations:
\begin{eqnarray}
\left[ \tilde{\beta}_{\bar{v}}(\bar{v},\bar{w})\frac{\partial}{\partial {\bar{v}}}+\tilde{\beta}_{\bar{w}}(\bar{v},\bar{w}) \frac{\partial}{\partial{\bar{w}}}+ \tilde{\eta}_t(\bar{v},\bar{w}) \right] \tilde{Z}_t^{-1}(\bar{v},\bar{w})=0
\end{eqnarray}
\begin{eqnarray}
\left[ \tilde{\beta}_{\bar{v}}(\bar{v},\bar{w}) \frac{\partial}{\partial{\bar{v}}}+\tilde{\beta}_{\bar{w}}(\bar{v},\bar{w})\frac{\partial}{\partial{\bar{w}}}+ A(\bar{v},\bar{w}) \right]  Y(\bar{v},\bar{w}) = \nonumber \\
B(\bar{v},\bar{w}).
\label{eqa8} 
\end{eqnarray}

The first equation looks like eq.(\ref{eq5}) and it is simply the restriction of the evolution to the $u_0=0$ plane; we can repeat our general arguments, obtaining in particular the RG trajectory that connects the $I$ and $XY$ fixed points. Notice however that, since the condition $\tilde{Z}_t^{-1}(\bar{v},\bar{w})=0$ is independent of $\bar{u}$, it defines a surface in full parameter space, and the above discussion shows that the fixed points $C$ and $T$ must lay on this surface.

Once the functions $\tilde{Z}_t^{-1}$ and $Y(\bar{v},\bar{w})$ have been determined, it is easy in the large $N$ limit to reconstruct the full $Z_t^{-1}=0$ surface, which can be simply described by the above condition and by the function:
\begin{eqnarray}
\bar{u}(\bar{v},\bar{w})=\frac{1}{Y(\bar{v},\bar{w})}
\end{eqnarray}

Notice that, as a consequence of eq.(\ref{eqa8}), the function $\bar{u}(\bar{v},\bar{w})$ does not depend on the detailed form of the RG function $\tilde{\eta}_t$, as expected from our general arguments.

\begin{figure}[htb]
%\vspace{-0.5cm}
\centerline{\psfig{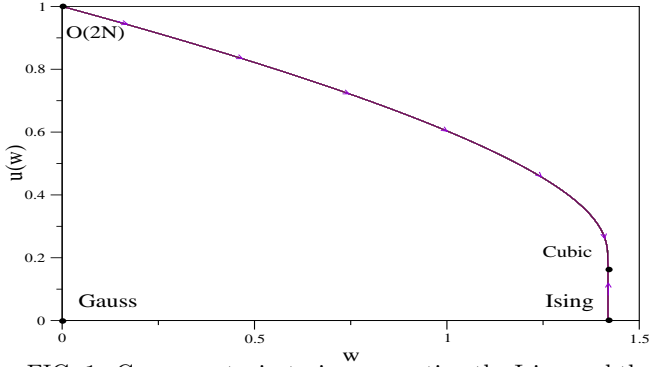}}

\caption{Crossover trajectories connecting the Ising and the $O(2N)$ fixed points to the Cubic one, in the limit of an infinite number of field components ($N \rightarrow \infty$).}
\label{rgflow}
\end{figure}
The intersections of the crossover surface with the planes $\bar{v}=0$ and $\bar{w}=0$ can now be found in a rather explicit form, by exploiting the above simplifications.

In terms of the generic variable $z$ we obtain the relevant equations:
\begin{eqnarray}
\left[ \tilde{\beta}(z) \frac{\partial}{\partial z}+\tilde{\eta}_t(z) \right] \tilde{Z}_t^{-1}(z)=0 
\end{eqnarray}
\begin{eqnarray}
\left[ \tilde{\beta}(z) \frac{\partial}{\partial z}+A(z) \right ]\left(\frac{1}{\bar{u}(z)} \right )=B(z)
\end{eqnarray}

It is straight-forward to solve the linear equations, obtaining
\begin{eqnarray}
\tilde{Z}_t^{-1}(z)= exp \left[- \int_{0}^{z} \frac{\tilde{\eta}_t(z')}{\tilde{\beta}(z')} dz' \right]
\end{eqnarray}
\begin{eqnarray}
 \bar{u}(z)=  \frac{X(z)}{\displaystyle \int_{0}^{z}\frac{B(z')}{\tilde{\beta}(z')}X(z')dz'}
\end{eqnarray}
\begin{eqnarray}
 \text{where} \quad X(z) = exp \left[\int_{0}^{z}\frac{A(z')}{\tilde{\beta}(z')}dz' \right]
\end{eqnarray}

It is easy to check that $\bar{u}(z)$ is a RG trajectory and that in the limits $z \rightarrow 0$, $z \rightarrow z^*$ ($\tilde{\beta}(z^*)=0$) we have $\bar{u}(0)= \frac{A(0)}{B(0)}$ and $\bar{u}(z^*)=  \frac{A(z^*)}{B(z^*)}$ respectively, consistent with the boundary conditions at the fixed points.

The above expressions lend themselves to simple analytical integration in the one-loop approximation and to easy numerical integration in the more general case.

For the sake of illustration we computed explicitly the crossover lines on the $(u,w)$ plane. Fig.\ref{rgflow} shows the results of our numerical integration of the equations, when resummed six-loop RG functions are employed. 

The straight line connecting $I$ to $C$ is the intersection of the $\bar{v}=0$ plane with the $\tilde{Z}_t^{-1}=0$ surface.

\section{An effective one-loop model}
It is possible to find a set of one-loop truncated RG equations which retains all the essential features of the above described model and lends itself to analytic integration, thus offering a rather explicit illustration of our results.

Our starting point is the set of equations:
\begin{eqnarray}
\beta_{u}=-u+a(v+w)u +u^2, \quad &&\beta_v= -v+v^2 \\
\beta_w=-w +2vw+2w^2, \qquad     &&\eta_t=-u-b(v+w)
\end{eqnarray}
where some trivial rescaling of the couplings is understood. 

The properties and symmetries of the fixed points are the same as in our general discussion, and in particular when $1\leq a \leq 2$ also the stability properties are exactly reproduced.
By direct integration one can find that:
\begin{eqnarray}
\tilde{Z}_t^{-1}(v,w)= (1-v)^{\frac{b}{2}}(1-v-2w)^{\frac{b}{2}}
\end{eqnarray}
and, being $F$ the standard hypergeometric function:
\begin{eqnarray}
&&Y(v,w)=  \\ &&(-1)^{\frac{a}{2}}w^{1-a}(v^2+2vw)^{\frac{a}{2}-1}\left[(1+\frac{v}{w})F\left (\frac{1}{2},\frac{a}{2},\frac{3}{2},(1+\frac{v}{w})^2 \right)\right. \nonumber \\
&&\left. -(1-2v+\frac{v}{w}(1-v))F\left(\frac{1}{2},\frac{a}{2},\frac{3}{2},(1-2v+\frac{v}{w}(1-v))^2\right)\right].\nonumber
\end{eqnarray}
One may easily check that the surface described by the function $u(v,w)=Y^{-1}(v,w)$ and the condition $\tilde{Z}_t^{-1}(v,w)=0$ enjoys all the properties discussed in the text.

% ========================= REFERENCES =========================


\begin{references}
\bibitem{B-B-l} C.~Bagnuls and  C.~Bervillier,
J. Phys. Lett. (Paris) {\bf 45},
L-95 (1984);
Phys.\ Rev.\ {\bf B 32}, 7209 (1985);
Phys.\ Rev.\ Lett.\ {\bf 58}, 435 (1987).


%\bibitem{B-B} C.~Bagnuls,  C.~Bervillier,
%Phys.\ Rev.\ {\bf B 32}, 7209 (1985);
%Phys.\ Rev.\ Lett.\ {\bf 58}, 435 (1987);
%\bibitem{B-B-prlf} C.~Bagnuls, C.~Bervillier,
%Phys.\ Rev.\ Lett.\ {\bf 58}, 435 (1987).


\bibitem{Fisher-prl} 
M.~E.~Fisher, Phys.\ Rev.\ Lett.\ {\bf 57}, 1911 (1986).


\bibitem{L-B-B-pre} 
E.~Luijten, H.~W.~J.~Bl\"ote and K.~Binder,
Phys.\ Rev.\ {\bf E 54}, 4626 (1996);
Phys.\ Rev.\ Lett.\ {\bf 79}, 561 (1997);
Phys.\ Rev.\ {\bf E 56}, 6540 (1997).


%\bibitem{L-B-B-prl} 
%E.~Luijten, H.~W.~J.~Bl\"ote and K.~Binder,
%Phys.\ Rev.\ Lett.\ {\bf 79}, 561 (1997);
%Phys.\ Rev.\ {\bf E 56}, 6540 (1997).



\bibitem{L1} 
E.~Luijten and K.~Binder,
Phys.\ Rev.\ {\bf E 58}, 4060 (1998);
Phys.\ Rev.\ {\bf E 59}, 7254 (1999);
Europhys.\ Lett.\ {\bf 47}, 311 (1999).







%\bibitem{L-B-B-pre} 
%E.~Luijten, H.~W.~J.~Bl\"ote and K.~Binder,
%Phys.\ Rev.\ {\bf E 54}, 4626 (1996).

%\bibitem{L-B-B-prl} 
%E.~Luijten, H.~W.~J.~Bl\"ote and K.~Binder,
%Phys.\ Rev.\ Lett.\ {\bf 79}, 561 (1997);
%Phys.\ Rev.\ {\bf E 56}, 6540 (1997).


\bibitem{prv} 
A.~Pelissetto, P.~Rossi and E.~Vicari,
Phys.\ Rev.\  {\bf E 58}, 7146 (1998);
Nucl.\ Phys.\ {\bf B 554}, 552 (1999).



\bibitem{Aharony-76} 
A.~Aharony, in
{\em Phase Transitions and Critical Phenomena},
edited by C.~Domb and J.~Lebowitz 
(Academic Press, New York, 1976),
Vol.\ 6, p. 357.



\bibitem{Pelissetto:2000ek}
A.~Pelissetto and E.~Vicari,
{\em ``Critical phenomena and renormalization-group theory''},
preprint IFUP-TH 2001-02, cond-mat/0012164,
and references therein.
%%CITATION = COND-MAT 0012164;%%

\end{references}
\end{document}